\documentstyle[psfig,epsfig,amstex,subfigure,afterpage,float]{mn}

\newcommand{\bu}{\mbox{\boldmath $u$}}

\newcommand{\ptl}{\partial}

\def\b0{\mbox{ {\bf 0}}}
\def\ltsima{\mbox{$\; \buildrel < \over \sim \;$}}
\def\simlt{\lower.5ex\hbox{\ltsima}}
\def\gtsima{\mbox{$\; \buildrel > \over \sim \;$}}
\def\simgt{\lower.5ex\hbox{\gtsima}}
\def\div{{\mathbf \nabla \cdot}}
\def\curl{{\mathbf \nabla \times}}
\def\grad{{\mathbf \nabla}}

\def\rc{r_{\rm c}}

\def\rin{r_{\rm in}}

\def\ur{u_r}
\def\ut{u_{\theta}}
\def\up{u_{\phi}}

\def\st{\sin\theta}

\def\s2t{\sin^2\theta}
\def\c2t{\cos^2\theta}
\def\sin{\mbox{  } {\rm sin}}
\def\cos{\mbox{  }{\rm cos}}

\def\Oc{\Omega_{\rm \star}}

\def\Oin{\Omega_{\rm in}}

\def\Enu{E_{\nu}}

\title{On rotationally driven meridional flows in stars}
\author[P. Garaud]{P. Garaud, \\  Department of Applied Mathematics and Theoretical Physics, University of Cambridge, Silver Street, CB39EW Cambridge, UK \\ Institute of Astronomy, University of Cambridge, Madingley Road, CB30HA Cambridge, UK \\ New Hall, Huntingdon Road, CB30DF Cambridge, UK}

\begin{document}
\maketitle

\begin{abstract}
A quasi-steady state model of the consequences of rotation on the hydrodynamical structure of a stellar radiative zone is derived, by studying in particular the role of centrifugal and baroclinic driving of meridional motions in angular-momentum transport. This nonlinear problem is solved numerically assuming
axisymmetry of the system, and within some limits, it is shown that
there exist simple analytical solutions. The limit of slow rotation recovers Eddington-Sweet theory, whereas it is shown that in the limit of rapid rotation, the system settles into a geostrophic equilibrium. The behaviour of the system is found to be controlled by one parameter only, linked to the Prantl number, the stratification and the rotation rate of the star. 
\end{abstract}

\section{Introduction}
\label{sec:intro}

The study of the effects of rotation on the hydrostatic structure of a star is a long-standing problem (Eddington, 1925, Sweet, 1950). It is the key towards a better understanding of stellar evolution, and consequently, of the whole observable universe. The effects of rotation on the stellar evolution can be
separated into two classes: effect on the contraction or expansion
with time of various stellar regions and rotational mixing (of chemical
species and angular momentum) through waves, instabilities and centrifugal
 driving of meridional motions. The effects of rotational mixing in stars was
reviewed by Pinsonneault (1997) and can be directly observed (e.g. Li
abundances and Li dip problem, MS turnoff, low-mass giants deep mixing
...). Recent works by Maeder \& Meynet (2000), or Chaboyer, Demarque \& Pinsonneault (1995) for example, attempt to consider the effects of rotational mixing on stellar evolution through the resolution of one-dimensional stellar models using various ad-hoc parameterizations. These models are successful in explaining some aspects of the observations but their predictive power is sometimes limited.

In this paper I propose to approach the problem through a new path,
which has different limitations and advantages thereby providing an
interesting complement to the existing models. I choose to study the
nonlinear effects of rotationally driven meridional flows on the
angular-momentum distribution of a stellar radiative zone. In the limit of 
a slowly rotating star, this is equivalent to the well-known Eddington-Sweet 
problem. In the case of rapid rotation, this problem has received less attention; it can now be studied through the formalism introduced in this paper. This type of approach has been considered in the past (e.g. Tassoul \& Tassoul (1983)); however, in previous works the nonlinearity of the angular-momentum transport processes was usually neglected mainly by assuming that the star was rotating nearly uniformly. As I shall show, this assumption is not self-consistent.

In Section 2, I present the equations for the hydrodynamic structure of a quasi-steady, laminar and axisymmetric radiative zone. The numerical results suggest a new scaling of the variables of the problem which can then be solved, in certain limits, analytically. The solutions are presented in Section 3. They provide the first self-consistent model of the effects of rotation on the centrifugal driving of meridional motions (assuming that the flow is a quasi-steady, laminar flow) and relies on one parameter only:
the mean rotation rate of the star. The results are discussed in Section 4.

\section{The model}
\label{sec:model}

I shall consider solar-type stars only and study their quasi-equilibrium 
structure under rotation. The overlying convection zone is affected by rotation mainly through the Coriolis distortion of the convective eddies. These effects are extremely complex; they are simply represented in this work through the rotation profile imposed by the convective zone to the underlying radiative zone. The radiative zone is assumed  to be a stable, isotropic fluid with uniform\footnote{This is a fairly good approximation; in any case, the viscosity only affects the dynamics of the system in thin boundary layers.} dynamical viscosity $\mu$.

The steady-state structure of the rotating radiative
zone is obtained by perturbing the hydrostatic equilibrium in the
following way:
\begin{align}
& \rho_{\rm h} \bu\cdot\grad\bu = -\grad \tilde{p}  - \rho_{\rm h}
\grad\tilde{\Phi} - \tilde{\rho} \grad \Phi_{\rm h} + \mu \grad^2 \bu
+ \frac{1}{3} \mu \grad(\div \bu) \mbox{   ,} \nonumber \\ & \rho_{\rm
h} T_{\rm h} \bu \cdot \grad s_{\rm h} = \div(k  \grad T) \mbox{   ,} 
 \frac{\tilde{p}}{p_{\rm h}}  =
\frac{\tilde{\rho}}{\rho_{\rm h}} + \frac{\tilde{T}}{T_{\rm h}} \mbox{
,}\nonumber  \\  & \grad^2 \tilde{\Phi} = 4\pi G \tilde{\rho} \mbox{
,   }  \div(\rho_{\rm h} \bu) = 0 \mbox{   .}
\label{eq:basic1}
\end{align}
where $p$ is the pressure, $T$ is the temperature, $\rho$ is the
density, $\Phi$ is the gravitational potential, $\bu = (\ur,\ut,\up)$
is the velocity field, $s$ is the entropy and $k$ is the thermal
conductivity. As the system is axisymmetric and near-uniform rotation 
is not assumed, there is no need to work 
in a rotating frame. Quantities denoted with suffix h are derived from the
non-rotating hydrostatic equilibrium equations, and perturbations to
that state are denoted with tildes. The equations are linearized with
respect to the background state (e.g. $\tilde{\Phi} \ll \Phi_{\rm h}$,
$\tilde{\rho} \ll \rho_{\rm h}$), but the full nonlinearity is kept in
the flow and in particular in the advection process
$\bu\cdot\grad\bu$. This approach is rigorously correct provided the
star is far from breakup (that is, when the mean ellipticity due to
rotation is much smaller than unity). The steady-state assumption is
valid provided the dynamical timescale of this system is short
compared to the stellar evolution and spin-down
timescales. This is not always necessarily the case, so that the direct
applicability of the results presented in Section 3 must be checked a
posteriori.

These equations must be supplemented with appropriate boundary
conditions.  Strictly speaking, correct boundary conditions would require a
theory of the rotation and circulation in the convection zone to which
the interior flows must be matched. 
Here, I shall simply assume that the forces maintaining 
the flow in the convection zone are so high that the angular velocity 
is unaffected by the flow in the underlying radiative zone, and is given by 
$\Omega_{\rm cz}(r,\theta) = \sum_{n=0}^{\infty} \Omega_n \cos^n \theta$
(where $\theta$ is the co-latitude and $r$ is the radial variable). 
In addition, I suppose that the stresses affecting the meridional circulation are viscous-like.  Since the effective viscosity in the convection zone is much greater than the viscosity in the radiative zone, it is not implausible that the continuity of stresses across the boundary imposes the constraints $\Omega = \Omega_{\rm cz}$, $\tau_{r,\theta} = r\ptl (\ut/r) /\ptl r + (1/r) \ptl \ur/\ptl \theta = 0$ and $\ptl \Omega/\ptl r$ is continuous (and therefore null) at the interface.  The inner core (for $r
< \rin$) is removed from the computational domain in order to avoid
singularities. This core is
impermeable and is assumed to rotate rigidly with angular velocity
$\Oin$ determined in such a way as to ensure a null angular-momentum
flux through the boundary. Finally, the regions outside the domain of
computation are assumed to satisfy $\grad^2 \tilde{T} = 0$ where
$\tilde{T}$ vanishes both at $r=0$ and when $r\rightarrow
\infty$ (since the inner core
supports no fluid motion and the effective
thermal conductivity in the convection zone is assumed to be much
larger than in the radiative zone).

Using the assumption of axisymmetry, I reduce the momentum equation in
(\ref{eq:basic1}) to:
\begin{eqnarray}
\bu \cdot \grad_{\xi} \left( \xi \sin\theta \up \right) &=& \Enu  {\rm D}^2 \left( \xi \sin \theta \up \right)
\mbox{   ,} \nonumber  \\  - \frac{1}{\rho_{\rm h}} \frac{\ptl}{\ptl
z} \left(\rho_{\rm h} \up^2 \right) &=& -  \frac{\sin\theta}{\rho_{\rm
h}} \left( \frac{\ptl \rho_{\rm h}}{\ptl \xi} \frac{\ptl
\tilde{\Phi}}{\ptl \theta} - \frac{1}{\epsilon} \frac{\ptl
\tilde{\rho}}{\ptl \theta} \right) \nonumber \\  &+& \Enu  {\rm D}^2 \left( \xi \sin\theta
\omega_{\phi} \right) \mbox{   ,}
\label{eq:basic2}
\end{eqnarray}
where $\bomega = \curl \bu$ is the vorticity, $\xi = r/\rc$ is the new
normalized radial coordinate, $z$ is the normalized cylindrical coordinate that
runs along the rotation axis, $\Enu = \nu/ \rc^2\Oc$ is the Ekman
number (where $\nu = \mu/\rho_{\rm h})$, 
and $\epsilon = \rc^2\Oc^2 (\ptl \Phi_{\rm h}/\ptl \xi)^{-1}$
is the ratio of the centrifugal to gravitational forces. The first of these equations simply describes angular momentum conservation, relating the advection term (LHS) to the diffusion term (RHS)\footnote{The diffusion term is expressed here in the
case of isotropic viscosity, but it is worth noting that any other
Reynolds-stress prescription for anisotropic transport can also be
used.}; the second equation is the azimuthal component of the curl of the momentum equation, commonly referred to in the geophysical literature as the `thermal wind equation'. In all these
expressions the following normalizations have been applied:
$\left[r\right] = \rc$, $ [u]=\rc\Oc$, $ [\tilde{\Phi}] = \rc^2
\Oc^2$, $[T] = 1K$, $[\rho] = 1$g/cm$^3$, where $\rc$ is the radius of
the radiative zone and $\Oc$ is the typical rotation rate of the
star. The operator $ {\rm D}^2$ is defined as
\begin{equation}
{\rm D}^2 = \frac{\ptl^2}{\ptl \xi^2} + \frac{\st}{\xi^2}
\frac{\ptl}{\ptl \theta} \left( \frac{1}{\st} \frac{\ptl }{\ptl
\theta} \right)
\label{eq:d}
\end{equation}

The energy equation becomes, to first order in the thermodynamical
perturbations
\begin{equation}
 T_{\rm h} \frac{\sigma N_{\rm h}^2}{\Oc^2} \frac{\epsilon}{\Enu} u_r  = \grad_{\xi}^2 \tilde{T}
\label{eq:energy}
\end{equation}
where $\sigma = \mu c_{\rm p} / k $ is the Prandtl number, and $N_{\rm
h}$ is the background buoyancy frequency.

Finally, the equation of state can be combined with the radial and
latitudinal components of the momentum equation to provide an
expression for $\tilde{\rho}$:
\begin{equation}
\frac{1}{\rho_{\rm h}} \frac{\ptl\tilde{\rho}}{\ptl \theta} =
\frac{\rho_{\rm h}}{p_{\rm h}}\rc^2 \Oc^2 \left[  \frac{\cos
\theta}{\sin\theta}\up^2 - \frac{\ptl \tilde{\Phi}}{\ptl \theta}
\right] - \frac{1}{T_{\rm h}} \frac{\ptl \tilde{T}}{\ptl \theta}
\label{eq:rho}
\end{equation}
This expression is used in the thermal wind relation in
(\ref{eq:basic2}) as well as in the latitudinal derivative of the
Poisson equation
\begin{equation}
\frac{\ptl }{\ptl \theta} \grad^2 \tilde{\Phi} = 4\pi G \left[
\frac{\rho^2_{\rm h}}{p_{\rm h}}\rc^2 \Oc^2 \left(  \frac{\cos
\theta}{\sin\theta}\up^2 - \frac{\ptl \tilde{\Phi}}{\ptl \theta}
\right) - \frac{\rho_{\rm h}}{T_{\rm h}} \frac{\ptl \tilde{T}}{\ptl
\theta} \right]
\label{eq:Poisson}
\end{equation}
The resulting equations (\ref{eq:basic2}), (\ref{eq:energy}) and
(\ref{eq:Poisson}) combined with the mass continuity equation are
solved for the unknowns $\ur,\ut,\up,\ptl \tilde{\Phi} / \ptl \theta$
and $\tilde{T}$.


\section{Numerical and analytical results}
\label{sec:anares}

The numerical study of the system presented in Section \ref{sec:model}
was performed using the standard solar model calculated by
Christensen-Dalsgaard et al. (1991) as the reference state. The
imposed latitudinal shear was chosen to be that observed in the solar
convection zone (with $\tilde{\Omega}_{\rm cz} = 1-0.15
\cos^2 \theta - 0.15 \cos^4 \theta$ in units of $\Omega_{\star}$).
 As the rotation rate $ \Omega_{\star}$ is varied, it
appears that only two parameters control the behaviour of the system:
the Ekman number (which controls the boundary-layer behaviour and the
flow velocities) and a new number 
\begin{equation}
\lambda = \sigma N_{\rm h}^2 /\Oc^2 \mbox{   ,}
\end{equation} 
which controls the dynamics of the bulk of the radiative
interior.

\subsection{ Slow rotation case: $\lambda \gg 1$}

In the case of slow rotation, the numerical results suggest that
 $\tilde{T}$ and the poloidal components of the velocity
 $u_{r,\theta}$ scale with $\Enu$ and $\lambda$ as:
\begin{equation}
\tilde{T} = \epsilon T_{\rm h} \overline{T} \mbox{   ,   }  u_{r,\theta} = \Enu/\lambda \overline{u}_{r,\theta}\mbox{   ,}
\label{eq:sca1}
\end{equation}
where the quantities with bars are the scaled quantities, of order of
unity. It is also found that $\up$ and $\tilde{\Phi}$ are always of order of
unity, which is expected. Note that the scaling for the meridional
motions is a local Eddington-Sweet scaling (see Spiegel \& Zahn,
1992). Using this ansatz into the system of equations given in
(\ref{eq:basic1}), an expansion in powers of $1/\lambda$ reveals that
the angular-momentum balance is dominated to zeroth order by viscous
transport through:
\begin{equation}
{\rm D}^2 (\xi\sin\theta \up) = 0 \mbox{   ,}
\label{eq:visc}
\end{equation}
which determines the angular velocity profile uniquely. Using this
result in the first order equations provides a relation between the
temperature and gravitational potential perturbations:
\begin{align}
& T_{\rm h} \frac{\ptl}{\ptl z} \left(\frac{1}{T_{\rm h}} \up^2
\right) = \sin\theta \left( \frac{\ptl \overline{T}}{\ptl \theta}
-\frac{\ptl \ln T_{\rm h}}{\ptl \xi} \frac{\ptl \tilde{\Phi}}{\ptl
\theta} \right) \mbox{   ,}  \\ &  \frac{\ptl}{\ptl \theta}
\grad_{\xi}^2 \tilde{\Phi} = - \frac{4 \pi G  \rho_{\rm h} \rc
}{g_{\rm h}} \left[ \frac{\ptl \ln p_{\rm h}}{\ptl \xi} \left(
\frac{\cos \theta}{\sin\theta}\up^2 - \frac{\ptl \tilde{\Phi}}{\ptl
\theta} \right) + \frac{\ptl \overline{T}}{\ptl \theta} \right] \mbox{
,} \nonumber
\label{eq:lb}
\end{align}
which can be solved independently for $\overline{T}$ and $\ptl
\tilde{\Phi}/\ptl \theta $. Finally, the temperature fluctuations lead
to meridional motions through the energy advection-diffusion equation:
\begin{equation}
\overline{u_r} = \grad^2_{\xi} \overline{T} \mbox{   .}
\label{eq:enb}
\end{equation} 

\begin{figure}
\centerline{\epsfig{file=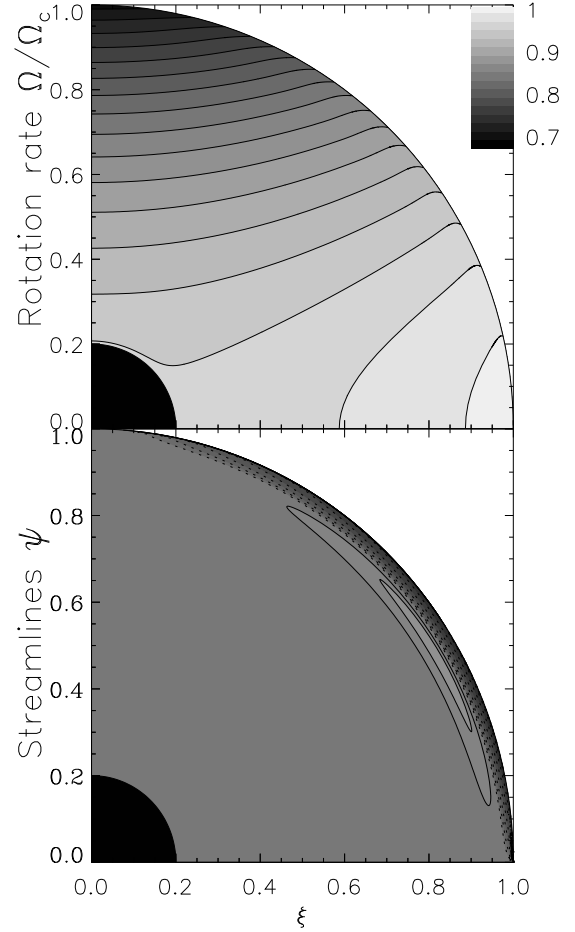,width=8cm}}
\vspace{-0.5cm}
\caption{\small Numerical solution for a solar-type star rotating 100
times slower than the sun ($\lambda \simeq 10^4$). The quadrants show
the radiative zone only. The upper panel shows the angular
velocity, which is viscously dominated. The interior rotation rate is
0.957 times that imposed at the surface at the equator. The lower
panel shows the streamlines (dotted lines represent a clockwise flow,
and solid lines represent an anti-clockwise flow). The contours are
logarithmically spaced.}
\label{fig:lambdabig}
\end{figure}

In Fig. \ref{fig:lambdabig}, I show the results of the numerical
solutions for the angular velocity profile and the meridional motions
corresponding to a slowly rotating solar-type star (for which $\lambda
\simeq 10^{4}$). The angular velocity profile is exactly the same as
one obtained through solving $(\grad^2 \bu)_{\phi} = 0$. The
meridional flow velocities decrease strongly with depth and the
streamlines follow a layered structure reminiscent of a Holton flow.

\subsection{Rapid rotation case: $\lambda \ll 1$}

\begin{figure}
\centerline{\epsfig{file=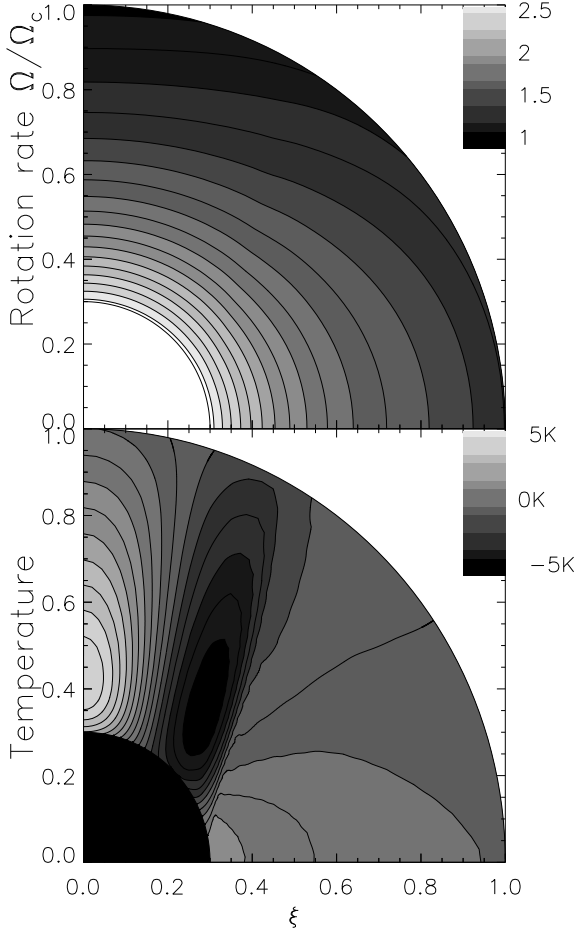,width=8cm}}
\vspace{-0.5cm}
\caption{\small Numerical solution of the system (\ref{eq:basic1}) for
a solar-type star rotating 10 times faster than the sun ($\lambda
\simeq 10^{-2}$). The upper panel shows the angular velocity, which varies strongly with depth and the lower panel shows the temperature fluctuations. 
Note that even when the stellar oblateness is of order
of $10^{-3}$, the temperature fluctuations remain of order of
$10^{-6}$.}
\label{fig:lambdasmall}
\end{figure}

\begin{figure}
\centerline{\epsfig{file=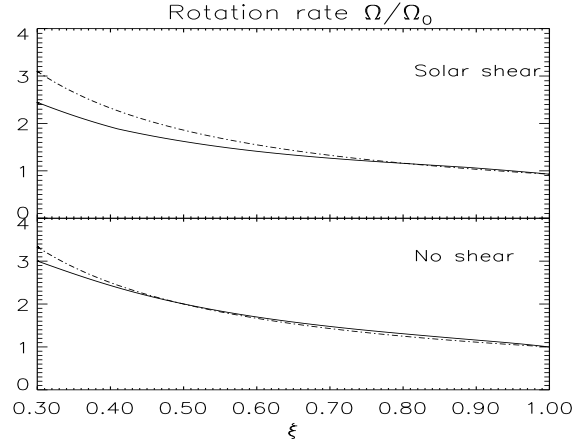,width=8cm,height=6cm}}
\caption{\small Comparison of the analytical prediction $\Omega(\xi)
\propto 1/\xi$ (dotted line) with the numerical results (solid line)
in the case of a solar shear (where the poles are rotating 30\% slower
than the equator) and in the case of no shear for a star with $\lambda \simeq 10^{-2}$. The no-shear case
satisfies the spherical approximation intrinsically well, so that the
analytical approximation is closer to the true solution in that case
than in the sheared case. The discrepancy with the true solution
increases near the core where the polytropic approximation to the
equation of state is no longer valid.}
\label{fig:log}
\end{figure}

In the case of rapid rotation the correct scaling seems to be
\begin{equation}
 \tilde{T} = \lambda \epsilon T_{\rm h} \overline{T} \mbox{   ,   } u_{r,\theta} = \Enu \overline{u}_{r,\theta} \mbox{   .}
\label{eq:sca2}
\end{equation}
This time, I perform an asymptotic expansion in the small parameter
$\lambda$. In this limit the temperature fluctuations are strongly
damped by the rapid heat diffusion (as $\lambda \ll 1$ is equivalent
to the small Prandlt number limit) and the system reaches an
equilibrium that is determined by the zeroth order equations:
\begin{eqnarray}
&& T_{\rm h} \frac{\ptl}{\ptl z} \left(\frac{1}{T_{\rm h}} \up^2
\right) = - \sin\theta \frac{\ptl \ln T_{\rm h}}{\ptl \xi} \frac{\ptl
\tilde{\Phi}}{\ptl \theta}  \mbox{   ,} \nonumber \\ &&
\frac{\ptl}{\ptl \theta} \grad_{\xi}^2 \tilde{\Phi} = 4\pi G
\frac{\rho^2_{\rm h}}{P_{\rm h}} \left[  \frac{\cos
\theta}{\sin\theta}\up^2 - \frac{\ptl \tilde{\Phi}}{\ptl \theta}
\right] \mbox{   .}
\label{eq:ls}
\end{eqnarray}
These equations describe a geostrophic equilibrium, and 
can in principle be solved for $\up^2$ and
$\tilde{\Phi}$ alone. The solutions provide, 
to the next order in $\lambda$, an equation for the
meridional flow through the advection diffusion balance:
\begin{equation}
\overline{\bu} \cdot \grad_{\xi} \left( \xi \sin\theta \up \right) =
{\rm D}^2 \left( \xi \sin \theta \up \right)\mbox{   ,}
\label{eq:ens}
\end{equation}
and, finally, the temperature fluctuations through
\begin{equation}
\overline{u}_r = \grad_{\xi}^2 \overline{T}\mbox{   .}
\label{eq:bla}
\end{equation}
 The results of the numerical simulations for small $\lambda$ ($\lambda
\simeq 10^{-2}$) are shown in Fig. \ref{fig:lambdasmall}.
It is interesting to note that the radial shear is much larger than
the latitudinal shear, which suggests to consider $\ptl \Omega/\ptl
\xi \gg \ptl \Omega /\ptl \theta$ to perform a first analytical approximation.
 Similarly,
I suppose that if $\Psi=\ptl \tilde{\Phi}/ \ptl \theta$ then $\ptl
\Psi/\ptl \xi \gg \ptl \Psi/\ptl \theta$. These approximations can be
used into equations (\ref{eq:basic2}) and (\ref{eq:Poisson}) to yield
the system
\begin{eqnarray}
&& \frac{1}{\xi^2} \frac{\ptl}{\ptl \xi} \left( \xi^2 \frac{\ptl
\Psi}{\ptl \xi}\right) = - \frac{4\pi G \rc^2 \rho^2_{\rm h} }{p_{\rm
h}} \left( \Psi + K\right) \mbox{   , } \nonumber \\ && \frac{\ptl K
}{\ptl \xi} = \frac{\ptl \ln T_{\rm h}}{ \ptl \xi} \left( \Psi + K
\right) \mbox{   , }
\label{eq:KPsi}
\end{eqnarray} 
where $K = \cos\theta \xi^2 \Omega^2$, which can easily be solved numerically
for $K$ and $\Psi$. However, if the radiative zone is crudely
approximated by a polytrope of index 3 then $T_{\rm h} \propto 1/\xi$ and $\rho_{\rm h} \propto 1/\xi^3$ between $\xi=0.3$ and $\xi = 1$. This leads to 
\begin{equation}
\frac{4\pi G \rc^2 \rho^2_{\rm h} }{p_{\rm h}} \simeq
\frac{\alpha_0}{\xi^2} \mbox{   and    } \frac{\ptl \ln T_{\rm h}}{
\ptl \xi} \simeq \frac{1}{\xi} \mbox{   . }
\end{equation}
These expressions can be used in the system (\ref{eq:KPsi}) to provide
a homogeneous relation between $\Psi$ and $K$, which can (amongst
other power law/oscillatory solutions) be solved analytically with
$\Psi \simeq  -K \simeq$ constant; this corresponds to $\Omega(\xi)
\propto 1/\xi$. The comparison of this very simple analytical
prediction with the results of the fully nonlinear numerical simulation
is presented in Fig. \ref{fig:log}, in the case where the convection
zone shear is null and in the case where it is of order of 30\% of the
mean rotation rate. When the imposed shear is small, the analytical prediction 
provides a good approximation to the numerical results.

\subsection{Lower boundary problem}

The numerical solutions for the angular velocity and the gravitational-field
 perturbation know little of the presence of a lower boundary; the
 meridional motions and the temperature fluctuations on the other hand
 seem to be strongly linked with the lower boundary conditions and may not
 represent what can be expected of a stellar radiative zone. In
 particular, the numerical simulations reveal the presence of an
 internal shear layer (related to the internal Stewardson layer in the
 Proudman spheres problem (Stewardson, 1966)), which is related to 
the presence of the impermeable lower boundary (and
 therefore mostly irrelevant to stellar hydrodynamics). This shear layer is
 reflected also in the temperature distribution (see Fig. \ref{fig:lambdasmall}).  However,
 as the meridional flow and temperature distribution affect the
 angular velocity profile only to the next order in $\lambda$, it is
 possible to calculate the flow pattern {\it as if there was no lower
 boundary} using the simple ansatz $\Omega \propto 1/\xi$ into
 equation (\ref{eq:ens}). Using the polytropic approximation again one can
 derive
\begin{equation}
\overline{\psi}  =  \xi \sin^2\theta \left[ \ln \left(\frac{
1-\cos\theta}{1+\cos\theta}\right) \right]  \mbox{   , }
\label{eq:stream}
\end{equation}
where $\overline{\psi}$ is the scaled stream-function defined as
$\overline{\bu} = - \curl( \overline{\psi}/\xi\sin\theta)$. 
This result suggests that in stars the dimensional
radial component of the meridional flow varies as $\ur \propto
\nu/ r $ independently of the stellar rotation rate when $\lambda \ll 1$. Note that this scaling is compatible with the results of the work of Talon et al. (1997). As Busse (1981) claimed that rapidly rotating systems settle down to purely zonal flows in the {\it inviscid} case, it is not surprising to find that the typical timescale for steady circulation in this rapid-rotation limit is the very slow viscous timescale. 

\section{Discussion and conclusions} 

In this paper I analysed the consequences of rotation on the driving of
meridional motions in stellar radiative zones. For a slowly rotating
star, this problem has been studied by Eddington (1925) and Sweet
(1950) who established that meridional motions are driven by 
stellar baroclinicity related to the rotation. For rapidly rotating
stars, the problem has never been solved self-consistently. I have
constructed a quasi-steady model of the rotating radiative zone of a 
solar-type (main-sequence)
star which can be solved numerically for a wide range of rotation
rates; through this numerical analysis I have been able to show that
the effect of rotation on the stellar structure can be described with
two parameters only: $\lambda = \sigma N_{\rm h}^2/\Oc^2$ and $\Enu =
\nu/\rc^2 \Oc$. The striking  reduction of this complicated nonlinear
problem to a rather simple form stems from the very close thermal
advection-diffusion balance; this is intrinsically related to the very
small Prandtl number typical of stellar radiative zones. In
comparison, studies of the effect of rotation on stratified shear
flows in the high-Prandtl-number limit reveal a very different
behaviour (e.g. Clark et al., 1971).

The system approaches asymptotic limits in the case of fast
rotation, $\lambda \ll 1$, and slow rotation, $\lambda \gg 1$. In both
cases, interesting analytical reductions of the problem can be used.
In the limit of slow rotation it was shown that
typical flow velocities scale as the local Eddington-Sweet velocity,
and the typical adjustment time of the system to the
steady state is an Eddington-Sweet timescale (as expected). As a result, for
molecular viscosities typical of solar-type stars it is unlikely that
the system would ever relax to the steady state found here. 

In the limit of rapid rotation ($\lambda \ll 1$), it was shown that the
system settles into a geostrophic equilibrium; this can be achieved on a
rather short timescale, which is linked with the generation and
dissipation of inertial waves. As a result, this type of equilibrium
is likely to be present in stars\footnote{although only if magnetic
fields and turbulent motions are too weak to influence that
equilibrium}. The resulting rotation profile varies with radius
roughly as $1/r$; the precision of this approximation depends 
on the degree of shear imposed by the
convection zone. Assuming this simple profile for the angular
velocity, the meridional motions can be deduced from angular-momentum
conservation. The typical
flow velocities {\it in the steady-state} are of order of $\nu/\rc$,
and are too slow to have any effect on material mixing. However, the 
continuous adjustment of the radiative zone to the stellar spin-down 
may lead to some significant mixing near the interface with the convection 
zone through the generation and dissipation of inertial waves. 
This can be studied self-consistently only through a time-dependent 
analysis of the flow, which is the subject of future work.

Of course, this analysis relies strongly on the fact that no 
dynamical effect save the rotation is present in the problem. The
major corrections to take into account would be turbulent motion, magnetic
field and chemical gradient. In regions where turbulent motions are
expected (i.e. possibly near the edge of a convective zone, or in
regions of strong shear) their effect could in principle be included
in equation (\ref{eq:basic1}), provided a Reynolds-stress tensor for
the microscopic motions is known. However, it is likely that the
turbulent Prandtl number is much closer to unity, which would then
render the analytical analysis obsolete (but not the numerical
abilities of the code). If magnetic fields are present, either on
large or small lengthscales, the whole dynamical structure of the star
may be changed completely. A good example of this effect is well
illustrated in the sun; observations
reveal that the radiative interior is rotating nearly uniformly, as the
shear imposed by the convection zone is suppressed within a very thin
layer (the tachocline). Models exist (e.g. Gough \& McIntyre, 1998) 
in which a small-amplitude, large-scale,
fossil magnetic field within the radiative zone is able to impose 
uniform rotation throughout most of the interior. This type of fossil
field is not unrealistic in any star. Finally, gradients in the mean
molecular weight may also significantly alter the dynamics of the
system by reducing or augmenting the effective buoyancy of the fluid,
especially near boundaries with convective regions. Mestel
(1953) has described how this effect may counterbalance the slow,
rotationally driven motions, and Theado \& Vauclair (2001) have argued that
this effect may be essential in explaining the observed Li abundances
in stars. 

As a result, the rotation profile in rapidly rotating stars may not
resemble that predicted in Section \ref{sec:anares}, but any observed
discrepancy with the predicted model will be a {\it clear signature}
of the effects discussed above.

\section*{Acknoledgments}

I thank Professors D. O. Gough, L. Mestel and J.-P. Zahn for many
fruitful discussions on this subject. I thank PPARC for financial support.

\end{document}